\documentclass[superscriptaddress,nobibnotes,amsmath,amssymb,notitlepage,twocolumn,pra,longbibliography]{revtex4-2}

\usepackage{bm,braket}
\usepackage[toc,page]{appendix}
\usepackage{comment}
\usepackage{dcolumn}

\usepackage{amsfonts}

\usepackage{graphicx,color,hyperref}
\usepackage[caption=false]{subfig}
\hypersetup{colorlinks=true, linkcolor=blue, citecolor=blue, urlcolor=blue} 

\graphicspath{{./img/}}

\newcommand{\beq}[1]{\begin{equation}\label{#1}}
\newcommand{\eep}{\;.\end{equation}}
\newcommand{\eec}{\;,\end{equation}}
\newcommand{\eeq}{\end{equation}}

\newcommand*\dd{\mathop{}\!\mathrm{d}} 

\newcommand{\ep}{\epsilon}

\newcommand{\om}{\omega}

\newcommand{\Om}{\Omega}

\DeclareMathAlphabet{\mathcal}{OMS}{cmsy}{m}{n} 

\renewcommand{\vec}[1]{{\bf #1}}

\newcommand{\kv}{\vec{k}}

\usepackage{amsmath}
\usepackage{amssymb}
\usepackage{xcolor}
\usepackage{bbm}
\usepackage{physics}
\usepackage{float}
\usepackage{graphicx}
\usepackage{dcolumn} 
\usepackage{bm} 
\usepackage{siunitx}
\usepackage{enumitem}  

\makeatletter
\renewcommand*{\fnum@figure}{{\normalfont\bfseries \figurename~\thefigure}}
\makeatother

\allowdisplaybreaks

\definecolor{orange}{rgb}{1,0.5,0}

\newcommand{\sect}[1]{\vspace{0.3em}{\it #1.}---}

\graphicspath{{./img/}}

\DeclareMathAlphabet{\mathcal}{OMS}{cmsy}{m}{n} 

\newcommand{\intBZ}{\int_{\text{BZ}}} 

\newcommand{\ii}{\mathrm{i}}

\makeatletter
\newcommand{\specificthanks}[1]{\@fnsymbol{#1}}
\makeatother

\begin{document}

\preprint{APS/123-QED}

\title{Magnetononlinear Hall effect from multigap topology in metal-organic frameworks}

\author{Chun Wang Chau}
\email{cwc61@cam.ac.uk}
\thanks{These authors contributed equally.}
\affiliation{TCM Group, Cavendish Laboratory, Department of Physics, J J Thomson Avenue, Cambridge CB3 0HE, United Kingdom}

\author{Wojciech J. Jankowski}
\email{wjj25@cam.ac.uk}
\thanks{These authors contributed equally.}
\affiliation{TCM Group, Cavendish Laboratory, Department of Physics, J J Thomson Avenue, Cambridge CB3 0HE, United Kingdom}

\author{Bo Peng}
\email{bp432@cam.ac.uk}
\thanks{These authors contributed equally.}
\affiliation{TCM Group, Cavendish Laboratory, Department of Physics, J J Thomson Avenue, Cambridge CB3 0HE, United Kingdom}

\author{Robert-Jan Slager}
\email{robert-jan.slager@manchester.ac.uk}
\affiliation{Department of Physics and Astronomy, The University of Manchester, Manchester M13 9PL, United Kingdom}
\affiliation{TCM Group, Cavendish Laboratory, Department of Physics, J J Thomson Avenue, Cambridge CB3 0HE, United Kingdom}

\date{\today}

\begin{abstract}
    We unveil that non-Abelian multigap band topology characterized by nontrivial Euler class \mbox{invariants} induces observable magnetononlinear Hall transport phenomena. We demonstrate these effects in a highly-tunable class of recently synthesized two-dimensional kagome N-heterocyclic \mbox{carbene} (NHC) metal-organic frameworks. We showcase the controllability of the nonlinear effect upon applying external voltage, changing temperature, and chemical substitutions that preserve the bulk topology and associated edge states. Our findings therefore reveal an uncharted presence of Euler class topology in metal-organic materials that can be experimentally deduced through measurable magnetotransport.
\end{abstract} 

\maketitle

\sect{Introduction} Quantum materials and phenomena host promising potential for novel fundamental effects and applications. Recent decades have seen an increasingly prominent role for the concept of topology in this discourse, due to its natural relationship with robust exotic phenomenological features~\cite{Rmp1,Rmp2,Rmp3}. Hallmark examples include integer and fractional quantum Hall effects \cite{vonKlitzing1980, Tsui1982, Laughlin1983, vonKlitzing1986}, and topological superconductivity~\cite{Kitaev2001, sato2017topological}, offering avenues for fault-tolerant quantum computation~\cite{Kitaev2003, Nayak2008,beenakker2013search}. With the extensive classifications of singly-gapped free fermion phases~\cite{Schnyder2008, Kitaev2009, Slager2013, Shiozaki2014, Kruthoff2017, Clas4, Clas5}, new developments have shown that uncharted multigap topological phases may emerge~\cite{bouhon2020a, Zhao_PT,Ahn2019a, Davoyan2024, Jankowski2024PRL, Jankowski2024PRB, Bouhon2020, BJY_linking, Lim2023_realhopf, Brouwer}, i.e., electronic band topologies associated with combinations of bands admitting more than a single spectral gap and multiple topological boundary state branches. While in metamaterials and quantum simulators multigap topologies are increasingly being explored~\cite{guo2021, Jiang2021, zhao2022observation, Jiang2024, Slager2024, Yang2024, Hu2024, Liu2025, Guillot2026,  Hu2026, jiang6bands}, and recently, quench, optical, and transport signatures were proposed~\cite{Uenal2020, Jankowski2025PRBoptical, Jain2025}, tunable material platforms and unique novel effects are still a subject of intense research efforts and pursuits. 

On a seemingly different note, organic materials are at the forefront of semiconductor industry~\cite{Forrest2007}, optoelectronic applications~\cite{Oksana2016}, and provide highly-tunable platforms for realizing exotic quantum phenomena~\cite{ssh1, ssh2}.  In this context, a remarkable controllability of structures, electronic states, and properties ranging from chemical catalysis to energy storage, can be achieved in metal-organic frameworks~\cite{Hiroyasu2013}. Structures of organic materials, in particular, offer a high tunability and control over electron-electron interactions which are vital for stabilizing unconventional band topologies~\cite{Po2019, Ahn2019a, Mondal2026} that otherwise remain unstable against strong correlations.

\begin{figure*}
\centering
\includegraphics[width=\linewidth]{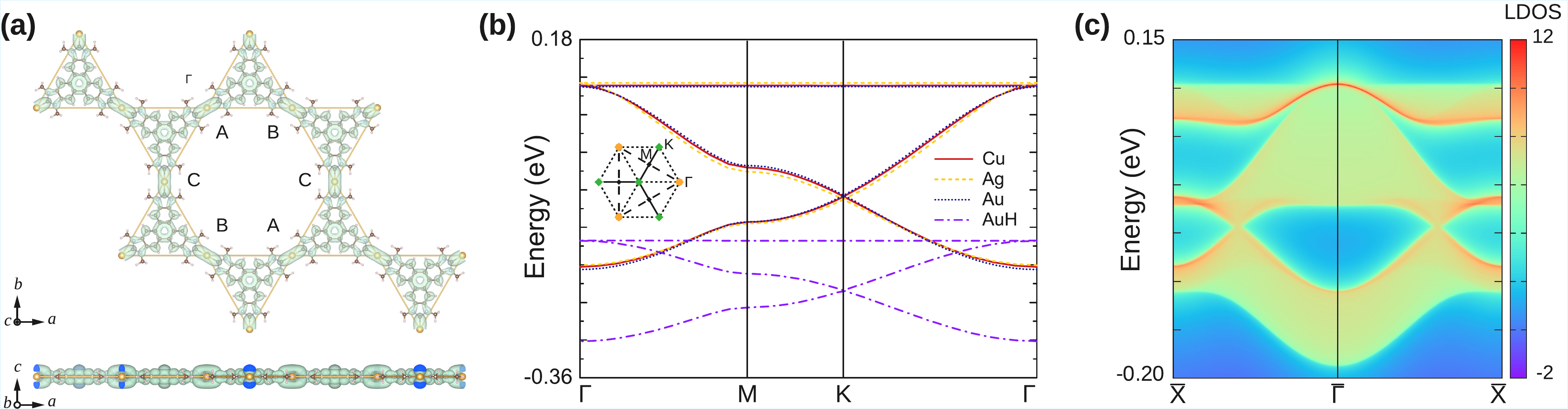}
  \caption{{\bf Multigap topology in metal-organic frameworks.}
  {\bf (a)} Crystal structure of two-dimensional metal-organic NHC-Au kagome lattice framework, top view (\textit{above}) and front view (\textit{below}). The partial charge density of the kagome bands near the Fermi level is shown by the light cyan isosurface, indicating delocalized features across the kagome lattice, against the localized orbitals around metallic atoms. The hybridized states of the gold atoms and NHC ligands constitute the kagome basis orbitals A,\,B,\,C. {\bf (b)} Comparison of band structure sections of kagome NHC-X ($\text{X} = \text{Au}, \text{Ag}, \text{Cu}, \text{AuH}$) metal-organic frameworks across hexagonal BZ, with quadratic band touching at the $\Gamma$ point (BZ center), and non-Abelian quaternion-charged Dirac nodes at the $\text{K}$ points (BZ corners). {\bf (c)} Multigap edge states from the Euler and quaternion-charged Dirac nodes (\textit{red}). The lower gap hosts edge states that are a consequence of $\pi$-Berry phase due to linear nodes with non-Abelian quaternion charges $q = \pm \text{i}$. The~upper gap hosts edge states from a topological quadratic node with patch Euler class $e_2 = 1$. 
  }
\label{Fig:Intr}
\end{figure*}

Topological matter and its unconventional properties can be captured by quantum anomalies realized through topological field-theoretic terms known from high-energy physics~\cite{Witten1988TQFT, Witten1995}. Microscopically, these can be understood as emergent from the topologically nontrivial wavefunctions and their responses~\cite{Qi2008}. Depending on the chemistry of the material, different topologies of electronic quantum states can be engineered in a subtle \mbox{interplay} with fundamental and crystalline symmetries \cite{Kruthoff2017, Clas4, Clas5}, and organic materials are no exception to that guiding rule. Recently, it was shown that tunable organic semiconductors not only realize topologically nontrivial states of electrons~\cite{Cirera2020}, but also of excitons~\cite{Jankowski2025exciton}, resulting in unusually enhanced exciton transport properties~\cite{Thompson2024exciton}. Given the high-tunability of organic materials, unusual transport under distinct external conditions, as induced by nontrivial topology of the quantum states realized therein, is expected to be achievable, but nevertheless has not been extensively studied to date~\cite{Pan2022SW, Wang2023o}.

In this work, we combine these progresses and uncover a~novel transport effect associated with the multigap topology, as well as a suitable material platform. Specifically, we retrieve an exotic quantum nonlinear magnetotransport in organic N-heterocyclic carbene (NHC) metal-organic framework materials and also discover the presence of an exotic type of non-Abelian electronic topology therein, mimicking the disclination structures of biaxial nematics ~\cite{Wu2019a, Bouhon2020}. These materials were recently synthesized under experimental conditions and their properties are an open platform for further \mbox{exploration}~\cite{Qie2024}. We show that these novel systems can realize topologically-induced anomalous magnetononlinear Hall transport, with electric current densities scaling bilinearly in electric and magnetic fields. Using first-principles calculations and analytical techniques, we demonstrate that the enhancement of such nonlinear currents is a more general consequence, and can serve as a~smoking-gun probe, of the non-Abelian topological states present in any two-dimensional electronic systems. Moreover, we show that the non-Abelian invariant in these NHC metal-organic frameworks, topologically quantified by the Euler characteristic class $(e_2\in \mathbb{Z})$, can~be related to the non-Abelian quaternion charges $(q\in \mathbb{H})$ of the band crossings emergent in a~multigap setup of the band structures of those materials. As~a~consequence, we~demonstrate that the presence of non-Abelian multigap nodal topologies results in multigap edge states in these systems.

\sect{Non-Abelian multigap topology}
We now discuss the non-Abelian multigap band topology, which we retrieve in the NHC-based kagome materials, see Fig.~\ref{Fig:Intr}(a). In~the three-band subspace realized close to the Fermi level, see Fig.~\ref{Fig:Intr}(b), the top two bands host a non-Abelian Euler invariant $e_2$ protected by the $\mathcal{C}_2 \mathcal{T}$, i.e., two-fold rotation combined with time-reversal, symmetry of the Hamiltonian, where the nodes between both pairs of bars realize band nodes with non-Abelian quaternion frame charges $q$: ${\mathbb{H} = \{\pm \text{i}, \pm \text{j}, \pm \text{k}, \pm 1 \}}$~\cite{Wu2019a}. Mathematically quaternionic nodes emerge as homotopy charges of a classifying space $\pi_1 [\mathsf{Fl}_{1,1,1}] \cong \mathbb{H}$ of multiply-gapped $\mathcal{C}_2 \mathcal{T}$-symmetric three bands, classified by a real flag variety ${\mathsf{Fl}_{1,1,1} \cong \mathsf{O}(3)/\mathbb{Z}_2^3}$~\cite{Jiang2021}. Moreover, such band nodes realize $\pi$-Berry phases leading to the presence of nodal topology-induced edge modes, see Fig.~\ref{Fig:Intr}(c). Merging and splitting of the band nodes satisfies the quaternion algebra, e.g., the quadratic band crossing with charge $q = -1$ can be split into two nodes of the same chirality $q = \text{i}$, on breaking the $\mathcal{C}_6$ symmetry of the kagome lattice~\cite{Jiang2021}.

The symmetry-protected stability of the quadratic band touching can be captured by the Euler class invariant. The patch Euler class $e_2$ is defined over a patch of the Brillouin zone (BZ), $\mathcal{D} \in \text{BZ}$, as, 
\beq{}\label{Eq:Euler}
    e_2 = \frac{1}{2\pi} \int_{\mathcal{D}} \dd^2 \kv~\mathrm{Eu} - \frac{1}{2\pi} \int_{\mathcal{\partial D}} \dd \kv \cdot \vb{a},
\eeq
where $\vb{a} = \bra{u_{n,\kv}} \ket{\nabla_\kv u_{n+1,\kv}}$ is the Euler connection and $\text{Eu} = \nabla_\kv \times \vb{a}$ is the Euler curvature between consecutive Bloch states $\ket{u_{n,\kv}}, \ket{u_{n+1,\kv}}$ at momentum $\kv$~\cite{Ahn2019a, Bouhon2020}. By definition, $\vb{a}$ is encoded in the off-diagonal elements of the non-Abelian Berry connection, ${A^c_{nm} = \bra{u_{n,\kv}} \ket{\partial_{k_c} u_{m,\kv}}}$.

The quadratic band touching in the NHC lattices hosts  topological invariant $e_2 = 1$ as long as the patch excludes the additional nodes from the $\text{K}$ points, see Fig.~\ref{Fig:Intr}(b). Moreover, the nodes at the $\text{K}$ points in the bottom gap host non-Abelian quaternion charges $q = \pm \text{i}$, which, on the contrary, can be gapped while the $\mathcal{C}_2 \mathcal{T}$ symmetry is preserved, similarly to the pairs of Dirac band nodes carrying $\pi$-Berry phases in graphene.

\begin{figure*}
\centering
\includegraphics[width=\linewidth]{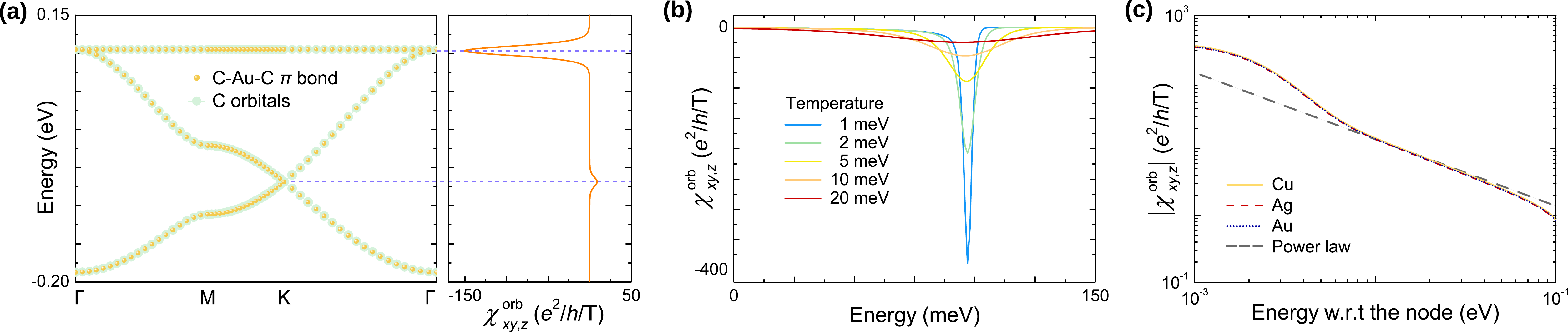}
  \caption{{\bf Magnetononlinear Hall effect from non-Abelian Euler band topology.} {\bf (a)} Orbital-projected kagome bands with $e_2 = 1$ in NHC-Au metal-organic framework and nonlinear magnetoelectric Hall response at different dopings~($\mu$). {\bf (b)}~\mbox{Temperature} dependence of the magnetononlinear Hall effect around the quadratic node in NHC-Au metal-organic framework. {\bf (c)}  Scaling of the magnetononlinear Hall response at temperature $T=1\;\mathrm{meV}$ in the NHC-X  ($\text{X} = \text{Au}, \text{Ag}, \text{Cu}$)  metal-organic frameworks, against the power-law dependence, $\chi^\text{orb}_{xy,z}\propto |e_2|/\mu$, arising from the multiband topology via \mbox{non-Abelian} Berry connection vorticity.}
\label{Fig:mNLHE}
\end{figure*}

Crucially, we retrieve all of the discussed signatures of the non-Abelian real topology in the band structures and local density of states (LDOS) of the metal-organic NHC materials, obtained using density functional theory (DFT). We furthermore find the multigap edge states from first principles, which manifestly appear in LDOS, see Fig.~\ref{Fig:Intr}(c). Finally, and most importantly, we connect the non-Abelian topology to the magnetononlinear Hall transport signatures, which allow to probe the multigap topology encoded in the non-Abelian Berry connection, definitional to the Euler bands, Eq.~\eqref{Eq:Euler}, and present in the kagome NHC metal-organic frameworks.

Before we focus on particular material realizations and the magnetononlinear transport effects, we comment on the general interplay of multigap Euler invariant with the $\pi$-Berry phases associated with bulk polar charge displacements~\cite{Vanderbilt2018berry} that result in multigap edge states in the NHC metal-organic lattices, as shown in Fig.~\ref{Fig:Intr}(c). The edge states in the gap hosting linear nodes, see Fig.~\ref{Fig:Intr}(b), can be understood as emerging from the $\pi$-Berry phases and associated Dirac strings~\cite{Jiang2021}, similarly to the boundary spectrum found in graphene and Fermi arcs found in Weyl semimetals. We note that like in graphene, this two-band subspace hosts trivial Euler class $e_2 = 0$. On the other hand, the quadratic touching with $e_2 = 1$ can be viewed as a non-Abelian Berry connection vortex with a geometric phase of $\phi = 2\pi$ associated with a~full winding of the non-Abelian Berry connection around the nodes~\cite{Ahn2019a, Jiang2021}. As such, the quadratic node viewed as a~full vortex of non-Abelian Berry connection, with its vorticity being captured by the Euler class $e_2$, induces edge states similarly to the half-vortices with $\pi$-Berry phases and charges $q = \pm \text{i}$, yet in another gap.

\sect{Metal-organic material realizations} 
We here detail on the material realizations of multigap topology in the NHC metal-organic frameworks, following their recent synthesis from NHC ligands and gold (Au) ions~\,\cite{Qie2024}. 

The originally synthesized monolayer NHC-Au framework crystallizes in $P6/mmm$ space group (No.\,191), and geometry optimization obtains a lattice constant of 27.0746\,\AA. Each central gold atom is connected by two co-planar NHCs, forming a~kagome lattice [Fig.\,\ref{Fig:Intr}(a)]. The $p_z$ orbitals of the gold atom form a three-centre $\pi$ bond with the $p_z$ orbitals the two adjacent carbon atoms from the NHC ligands. Notably, the \mbox{C-Au-C} $\pi$ bond contributes dominantly to the kagome bands near the Fermi level. As shown in Fig.\,\ref{Fig:mNLHE}(a), the contribution from the C-Au-C $\pi$ bond to the kagome bands is nearly the same as one with the remaining 48 carbon atoms. The partial charge density of the kagome bands is shown by the light cyan isosurface in Fig.\,\ref{Fig:Intr}(a), showing the expected orbital basis-hybridization features. For further details on the first principles DFT simulations, see Methods.


Centrally to this work, the three kagome bands form two degenerate points at high-symmetry points. The upper two bands cross at $\Gamma$, and the lower two bands cross at K. Benefiting from the three kagome bands in the $P6/mmm$ kagome lattice with $\mathcal{C}_2 \mathcal{T}$ symmetry, the band crossing points formed by the three bands at $\Gamma$ realizes multigap topology with patch Euler class $e_2 = 1$. 


In terms of chemical tunability, substituting the gold atoms with copper (Cu) or silver (Ag) leads to \mbox{NHC-Cu} and NHC-Ag lattice with similar kagome bands [Fig.\,\ref{Fig:Intr}(b)], which is expected because of the analogous kagome structures. On the other hand, the adsorption of hydrogen atoms on the top site of the gold atom leads to NHC-AuH with a large band gap $>1$\,eV. The \mbox{C-AuH-C} bond forms the three kagome bands occupying the highest valence states, with larger contribution than all carbon orbitals. Given the analogous electronic structures within the broader class of NHC-X materials, as~shown in Fig.~\ref{Fig:Intr}(b), we retrieve the multigap edge states, c.f., Fig.~\ref{Fig:Intr}(c), and responses (see Fig.~\ref{Fig:mNLHE}) of~\mbox{NHC-Au}, \mbox{NHC-Ag}, \mbox{NHC-Cu} metal-organic frameworks. We find the responses (i) under different external voltages determining electrostatic doping through the associated chemical potential $(\mu)$, (ii) over a~range of temperatures $(T)$ allowing to control the magnitude of the effect through distinct thermal band occupations, while the Euler \mbox{invariant} $e_2$ is preserved.

In the following, we show how our central finding of the correspondence between the nontrivial multigap topology and magnetononlinear response is universally robust over a range of external physical conditions applied to distinct NHC-X complexes, promising for an experimentally measurable smoking-gun response feature of the Euler class. 

\sect{Magnetononlinear Hall transport}
We now focus on the universal details of the nonlinear magnetotransport response induced by non-Abelian Euler band topology, which we retrieve in the NHC-based metal-organic frameworks. The intrinsic response is of second order in the static electromagnetic fields.
The dominant second-order response of interest is given by the current densities $j_i$,
\beq{}
    j_i = \chi^\text{orb}_{ij,k} E_j B_k,
\eeq
with $i,j,k = x,y,z$, and $E_j$, $B_k$ the electric and magnetic field components. The non-Abelian Berry connection-induced orbital contribution to the second-order magnetoconductivity of a two-dimensional crystal is given~by~\cite{Gao2014, Wang2024_1, Wang2024_2}
\beq{}\label{Eq:Chi}
    \chi^\text{orb}_{ij,k} = \intBZ \frac{\dd^2 \kv}{(2\pi)^2}~\sum^{}_n \frac{\partial f_{n,\kv}}{\partial E_{n,\kv}} [v_j^{nn} F^{\text{O},n}_{ki} - v_i^{nn} F^{\text{O},n}_{kj}],
\eeq
with $E_{n,\textbf{k}}$ the band energies, $f_{n,\textbf{k}} = \Big[1+\text{exp}\big(\frac{E_{n,\textbf{k}} - \mu}{k_B T}\big)\Big]^{-1}$, the Fermi--Dirac occupation factors which depend on the chemical potential $\mu$ and temperature $T$, and ${v_i^{ml} = \partial_{k_i} E_{m,\kv} \delta_{ml} + (1-\delta_{ml}) A_i^{ml}/(E_{m,\kv}-E_{l,\kv})}$, the matrix elements of the electron velocity operator $v_i$. The anomalous magnetic orbital polarizability $F^{\text{O},n}_{kj}$ in band $n$ reads~\cite{Wang2024_1},
\beq{}
    F^{\text{O},n}_{kj} =  \sum_{m,l \neq n} \text{Re} \Big[ \varepsilon_{kpq} \frac{(v_q^{ml} + \delta_{lm} v_q^{nn}) A_p^{ln} A_j^{mn}}{E_{n,\kv}-E_{m,\kv}} + \frac{1}{2} \partial_{k_q} g^n_{pj} \Big],
\eeq
where $g^n_{ij} = \sum_{m \neq n} \text{Re}~( A_{i}^{nm} A_{j}^{mn})$ is the quantum metric realized in band $n$~\cite{Provost1980, Ahn2021, bouhon2023quantum, Resta_2011}. We observe that the effect is governed purely by the non-Abelian Berry connection $A^a_{nm}$, and the dispersion present in the bands $E_{n,\kv}$, which can be deduced from angle-resolved photoemission (ARPES) experiments. Therefore, the magnetononlinear Hall response allows to access the present non-Abelian band geometry captured by the Berry connection, as well as the underlying non-Abelian Euler band topology that induces such non-trivial geometry. 

We find that in the metal-organic frameworks central to this work, the in-plane current $j_x$ emerges in response directly proportional to the perpendicular in-plane electric field $E_y$ and perpendicular magnetic field $B_z$, as~governed by the magnetoconductivity tensor component $ \chi^\text{orb}_{xy,z}$ (see Fig.~\ref{Fig:mNLHE}) captured by the non-Abelian Berry connection elements $A_x^{nm}, A_y^{nm}$. Physically, the magnetic field induces both an orbital magnetization reflecting the nontrivial multiband topology~\cite{Chau2026}, and a field-induced Berry curvature~\cite{Gao2014} that supports Hall response under the additional electric field. Notably, the Berry curvature vanishes in the absence of magnetic field due to the $\mathcal{C}_2 \mathcal{T}$ symmetry. For more technical details, and the derivation of the response, see Methods. 

In Fig.~\ref{Fig:mNLHE}, we show the nonlinear Hall magnetoresponse at different dopings $(\mu)$ and temperatures $(T)$, which corresponds to different tunable occupations of non-Abelian band nodes in different gaps. The orbital part of the nonlinear magnetoconductivity tensor $\chi^\text{orb}_{xy,z}$~\cite{Liu2023,Liu2023a,Liu2023b}, is shown in the right bottom of Fig.\,\ref{Fig:mNLHE}(a). We observe a~strong response peak for nodes with Euler class $e_2 = 1$ at $\Gamma$~point in the BZ, which we further validate with effective tight-binding models (see Methods). The thermal stability of this peak is demonstrated in Fig.\,\ref{Fig:mNLHE}(b), and the universality of the scaling of the peak response, which we identify as $\chi^\text{orb}_{xy,z }\propto |e_2|/\mu$, is shown for multiple NHC metal-organic materials in Fig.\,\ref{Fig:mNLHE}(c). We provide an analytical derivation for the~universal magnetononlinear response scaling within effective continuum models of the Euler bands in the Supplementary~Information~(SI).

\sect{Discussion} Having retrieved the non-Abelian multigap topology in the NHC-based metal-organic framework materials, we now discuss the obtained results. We first note that the studied magnetononlinear Hall effect requires nontrivial multiband contributions from the non-Abelian Berry connection. Namely, the response vanishes if the multiband non-Abelian band geometry induced by the Euler topology is trivialized. Furthermore, the multigap edge states physically confirm the presence of non-Abelian charges and Euler invariants realized between different bands in the first-principles calculations, as~demonstrated in the projected LDOS.

Having unravelled the magnetononlinear Hall effects due to the non-Abelian multigap Euler topologies in the metal-organic kagome frameworks, multiple avenues for the applications arise. First, it should be noted that as a perturbative effect in magnetic fields, unlike integer quantum Hall effect, the anomalous transport retrieved in this work requires small magnetic fields. The identified anomalous magnetononlinear Hall effect results in dissipationless currents analogously to the linear Hall effects, which opens possibilities for applications in sensing devices operating at low magnetic fields. Furthermore, the~retrieved nonlinear dissipationless charge transport could be further adapted for nanoelectronic circuits that leverage the showcased exotic bulk and boundary multigap topological states.

In summary, we uncover non-Abelian multigap Euler band topology in organic materials, which can be directly probed through a nonlinear magnetotransport response. We show how the response emerges in the class of (NHC) carbene-based metal-organic crystals, demonstrating that it is a universal feature of the multigap Euler topology realized in such systems. We further confirm the presence of multigap topology by retrieving the non-Abelian charge induced edge states and demonstrate the robustness of such exotic features against electrostatic doping, temperature, and chemical modifications, which provides promises for realistic technological applications in electronic devices.

\sect{Acknowledgments} The authors correspondingly thank Steven G. Louie for providing crystal structure data, Xiaoxiong Liu for helpful discussions on nonlinear transport calculations, and Adrien Bouhon for numerous discussions on non-Abelian band topologies.  C.W.C. acknowledges funding from the Croucher Cambridge International Scholarship by the Croucher Foundation and the Cambridge Trust. W.J.J.~acknowledges funding from the Rod Smallwood Studentship at Trinity College, Cambridge. B.~P. acknowledges support from Magdalene College Cambridge for a Nevile Research Fellowship.  R.-J.S. acknowledges funding from an EPSRC ERC underwrite grant  EP/X025829/1. This research was supported in part by grant NSF PHY-2309135 to the Kavli Institute for Theoretical Physics~(KITP).


\begin{widetext}

\section*{Methods}

\subsection{First-principles calculations}

We perform first-principles calculations using the Vienna \textit{ab initio} Simulation Package ({\sc VASP})\,\cite{Kresse1996,Kresse1996a} with the projector-augmented wave (PAW) basis set\,\cite{Bloechl1994,Kresse1999}. We use the Perdew-Burke-Ernzerhof exchange-correlation functional revized for solids (PBEsol)\,\cite{Perdew2008} under the generalized gradient approximation (GGA) formalism. A plane-wave cutoff of 700\,eV is used with the Brillouin zone sampled by a $\mathbf{k}$-mesh of $3\times3$. The lattice constants and atomic coordinates are fully relaxed using with energy and force convergence criteria of $10^{-6}$\,eV and $10^{-2}$\,eV/\AA\ respectively. The interlayer spacing is larger than 21\,\AA, and dipole corrections are applied\,\cite{Makov1995} to avoid interactions between periodic neighbours along $z$. The inclusion of spin polarization and spin-orbit coupling confirms non-magnetic order. To analyze the topological properties, maximally localized Wannier functions\,\cite{Marzari1997,Souza2001,Marzari2012} for the $p_z$ orbitals of Au/Ag/Cu are generated using {\sc wannier90}\,\cite{Mostofi2008,Mostofi2014,Pizzi2020}. 
The edge states are computed using {\sc WannierTools}\,\cite{Wu2018}. The transport properties are computed from {\sc WannierBerri}\,\cite{Tsirkin2021} using the recently-implemented nonlinear transport formulations\,\cite{Liu2023,Liu2023a,Liu2023b}.

\subsection{Tight-binding Hamiltonians}

We generate the tight-binding Hamiltonians upon Wannierizing the DFT bands. The general Hamiltonian for a~class of the considered NHC metal-organic materials reads,
\beq{}
    H = \sum_{i} \varepsilon_i c^\dagger_{i} c_{i} -t \sum_{\langle i, j \rangle} c^\dagger_{i} c_{j} - t' \sum_{\langle \langle i, j \rangle \rangle} c^\dagger_{i} c_{j},
\eeq
where $i,j$ run over the kagome orbital flavours $A, B, C$ predominantly contributed by the \mbox{C-X-C} $\pi$ bond bridges, and where $t$ and $t'$ are nearest neighbor (NN) ($\langle i, j \rangle$) and next-nearest neighbor (NNN) ($\langle \langle i, j \rangle \rangle$) hoppings. The hoppings manifestly respect the $\mathcal{C}_6$ symmetry present in the materials. The corresponding tight-binding parameters obtained from the Wannierized DFT bands of distinct NHC-X metal-organic frameworks are listed in the Table below:

\begin{equation*}
    \begin{tabular}{|c|c|c|c|}
        \hline\
        NHC-X &
        $\ep_i$ (meV) & $t$  (meV)& $t'$  (meV)\\
        \hline
        Au & -4.6 & 50.8 & 50.8
        \\
        \hline
        Ag & -4.8 & 51.1 & 51.1
        \\
        \hline 
        Cu & -4.4 & 50.5 & 50.6
        \\
        \hline
    \end{tabular}.
\end{equation*}

\subsection{Magnetononlinear Hall effect}

We hereby include the derivation of the magnetononlinear Hall effect central to this work, following closely Ref.~\cite{Gao2014}. To~first order, the magnetic field $\vec{B}$ perturbs the Bloch states, $\ket{u_{n,\textbf{k}}} \rightarrow \ket{u_{n,\textbf{k}}} + |{u'_{n,\textbf{k}}} \rangle$, via a magnetic Hamiltonian coupling, $\Delta H = - \textbf{m} \cdot \vec{B}$, where $\textbf{m}$ is an orbital magnetization operator. The first-order eigenstate corrections $|{u'_{n,\textbf{k}}}\rangle$ read,
\beq{}
    \ket{u'_{n,\textbf{k}}} = \sum_{m \neq n} \frac{\bra{u_{m,\textbf{k}}} \Delta H \ket{u_{n,\textbf{k}}}}{E_{n,\textbf{k}}-E_{m,\textbf{k}}} \ket{u_{m,\textbf{k}}}.
\eeq
Correspondingly, the magnetic perturbation introduces a field-induced correction to the Berry connection,
\beq{}
    (A^{nn}_i)' = \ii \Big(\bra{u_{n,\textbf{k}}} \ket{\nabla_\kv u'_{n,\textbf{k}}} + \bra{u'_{n,\textbf{k}}} \ket{\nabla_\kv u_{n,\textbf{k}}} \Big) = F^{n}_{ij} B_j.
\eeq
where, to this end, we adapt the Einstein summation convention. On assuming that the spin susceptibility contributions are negligible compared to the dominant topologically-induced orbital contributions: $F^{n}_{ij} \approx F^{\text{O},n}_{ij}$, where $F^{\text{O},n}_{ij}$ reads, 
\beq{}
    F^{\text{O},n}_{ij} = \sum_{m \neq n} \frac{A_i^{mn} \om_j^{mn}} {E_{n,\textbf{k}}-E_{m,\textbf{k}}}.
\eeq
Here, the $\om^{mn}_i$ matrix elements account for the field-induced orbital magnetization and read
\beq{}
\om_j^{mn}=  -\ii \varepsilon_{jki} \sum_{l \neq n} \frac{(v_i^{ml} + \delta_{lm} v_i^{nn}) A_k^{ln}}{E_{n,\kv}-E_{m,\kv}}.
\eeq
The field-induced Berry curvature correction reads $(\mathbf{\Om}^n)' \equiv \nabla_\kv \times (\vec{A}^{nn})'$ and culminates in the magnetic field-dependent second-order current under the action of electric field $\textbf{E}$~\cite{Gao2014},
\beq{}
    \textbf{j}^{(2)} = \frac{e^2}{\hbar} \intBZ \frac{\dd^2 \textbf{k}}{(2\pi)^2} \sum_n f_{n, \textbf{k}} [\textbf{E} \times (\mathbf{\Om}^n)' ],
\eeq
\beq{}
    j^{(2)}_i = \frac{e^2}{\hbar} \intBZ \frac{\dd^2 \textbf{k}}{(2\pi)^2} \sum_n f_{n, \textbf{k}} \varepsilon_{ijk} \varepsilon_{kpq} E_j B_r ( \partial_{k_p} F^\text{O,n}_{qr}).
\eeq
Upon factoring out the electromagnetic fields $(E_j, B_k)$, and using an identity, $\varepsilon_{ijk} \varepsilon_{kpq} = \delta_{ip} \delta_{jq} - \delta_{iq} \delta_{jp}$, the~Eq.~\eqref{Eq:Chi} is obtained on integration by parts with respect to the momentum space derivatives, $\frac{1}{\hbar} (\partial_{k_p} f_{n,\textbf{k}}) F^\text{O,n}_{qr} = \frac{1}{\hbar}\frac{\partial f_{n,\textbf{k}}}{\partial E_{n,\textbf{k}}} (\partial_{k_p} E_{n,\textbf{k}})F^\text{O,n}_{qr} = \frac{\partial f_{n,\textbf{k}}}{\partial E_{n,\textbf{k}}} (v^{nn}_pF^\text{O,n}_{qr})$, consistently with Ref.~\cite{Gao2014}. Notably, in the phases with the nontrivial Euler invariant, the field-free Berry curvature, $\mathbf{\Om}^n = 0$, vanishes identically under the $\mathcal{C}_2 \mathcal{T}$ symmetry~\cite{Bouhon2020} equivalent to the spatiotemporal inversion ($\mathcal{PT}$) symmetry in the considered two-dimensional systems. Therefore, as the magnetic field induces a field-dependent correction, $\mathbf{\Om}^n  \rightarrow \mathbf{\Om}^n + (\mathbf{\Om}^n)'= (\mathbf{\Om}^n)'$, the field-induced second-order current $\textbf{j}^{(2)}$ becomes the leading Hall current contribution, while the first-order Hall current (due to $\mathbf{\Om}^n$) vanishes.

\end{widetext}

\bibliography{references}

\end{document}


\title{SUPPLEMENTARY INFORMATION \\
Magnetononlinear Hall effect from multigap topology in metal-organic frameworks}

\author{Chun Wang Chau}
\email{cwc61@cam.ac.uk}
\thanks{These authors contributed equally.}
\affiliation{TCM Group, Cavendish Laboratory, Department of Physics, J J Thomson Avenue, Cambridge CB3 0HE, United Kingdom}

\author{Wojciech J. Jankowski}
\email{wjj25@cam.ac.uk}
\thanks{These authors contributed equally.}
\affiliation{TCM Group, Cavendish Laboratory, Department of Physics, J J Thomson Avenue, Cambridge CB3 0HE, United Kingdom}

\author{Bo Peng}
\email{bp432@cam.ac.uk}
\thanks{These authors contributed equally.}
\affiliation{TCM Group, Cavendish Laboratory, Department of Physics, J J Thomson Avenue, Cambridge CB3 0HE, United Kingdom}

\author{Robert-Jan Slager}
\email{robert-jan.slager@manchester.ac.uk}
\affiliation{Department of Physics and Astronomy, The University of Manchester, Manchester M13 9PL, United Kingdom}
\affiliation{TCM Group, Cavendish Laboratory, Department of Physics, J J Thomson Avenue, Cambridge CB3 0HE, United Kingdom}

\date{\today}
\maketitle

\section*{Magnetononlinear Hall response of Euler bands from a continuum model}

To describe the low-energy physics of a pair of Euler bands present in the NHC metal-organic frameworks central to the main text, we fit an effective two-band $\kv\cdot \mathbf{p}$ model governed by a Hamiltonian $H_\text{Eu}$:

\begin{align}
    H_{\text{Eu}} &= \frac{p^2}{2m_1}P_1 + \frac{p^2}{2m_2}P_2
    \nonumber\\
    &= \frac{1}{4m_1m_2}[(m_1+m_2)p^2\s_0
    +2(m_1-m_2)p_xp_y\s_1+(m_1-m_2)(p_x^2-p_y^2)\s_3]\;,\\
      P_1 & = \ket{u_1}\bra{u_1} =\begin{pmatrix}
        \sin^2\th & -\sin\th\cos\th
        \\
        -\sin\th\cos\th & \cos^2\th
    \end{pmatrix}\ ,
    \\
    P_2 &= \ket{u_2}\bra{u_2} = \begin{pmatrix}
        \cos^2\th & \sin\th\cos\th
        \\
        \sin\th\cos\th & \sin^2\th
    \end{pmatrix}\ ,
\end{align}
where we have defined $p_x=p\cos\th$ and $p_y=p\sin\th$, with dispersion $\ep_1=p^2/2m_1$ and $\ep_2=p^2/2m_2$ respectively.  

We now proceed to the computation of the magnetononlinear Hall response of the Euler bands effectively captured by the continuum Hamiltonian $H_\text{Eu}$. Following Ref.~\cite{Gao2014}, the field-induced positional shift of electrons in the lower band ($n=1$), in terms of the unperturbed Bloch bands, is~given upon an action of a pair of external fields $B_j, E_j$ by:
    \begin{equation}
        A_i' = F_{ij}B_j + G_{ij}E_j\; ,
    \end{equation}
    where the corresponding Berry connection polarizabilities read~\cite{Gao2014}:
    \begin{align}\label{F}
        F_{ij}
        &=
        \mathrm{Im}
        \sum_{n\neq 1}
        \frac{v_i^{1n}\om_j^{n1}}
        {(\ep_1-\ep_n)^2}\; ,
        \\
        \label{G}
        G_{ij}
        &=
        2\mathrm{Re}
        \sum_{n\neq 1}
        \frac{v_i^{1n} v_j^{n1}}
        {(\ep_1-\ep_n)^3}\; ,
    \end{align}
    with $\om^{1n}$ defined as:
    \begin{equation}
        \om_j^{1n}
        =
        -\text{i}\ep_{jkl}
        \sum_{m\neq 1}
        \frac{[v_k^{nm}+v^{nn}_k\d_{nm}]v_l^{m1}}
        {\ep_m-\ep_1}\; ,
    \end{equation}
    and $\ep_1$, $\ep_n$, the band dispersions. Moreover, in the above, one defines a velocity operator matrix $(v^{nm}_i)$:
    \begin{align}
        v_i^{nm}
        &=\bra{u_{n}}\partial_{p_i}H\ket{u_{m}}\; ,
    \end{align}
    where $\ket{u_m}$ is the Bloch state of a band $m$, for a Hamiltonian $H$. The considered two-band model Hamiltonian can be succintly decomposed as:
    \begin{equation}
        H_\text{Eu} = h_0+ \h\cdot\sv\; .
    \end{equation}
    For the lower band, following Ref.~\cite{Gao2014}, one obtains from Eqs.~\eqref{F} and \eqref{G}:
    \begin{align}
        F_{ij} &=
        -\frac{g_{ik}\varepsilon_{klj}\partial_{p_l}h_0}{4h}
        -\frac{1}{8}\varepsilon_{jkl}\G_{lki}\; ,
        \nonumber\\
        G_{ij} &=
        -\frac{1}{4h}g_{ij}\; ,
    \end{align}
    where the interband quantum metric is given by $g_{ij}=\partial_{p_i}\n\cdot\partial_{p_j}\n$, with $\n=\h/h,\; h=\vert\h\vert$. Here, we also adapt the Christoffel symbols, $\G_{lki}=(\partial_{p_i}g_{lk}+\partial_{p_k}g_{li}-\partial_{p_l}g_{ki})/2$.

As relevant to the main text, for $H_\text{Eu}$, we have:
\begin{align}
    h_0 &=
    \frac{m_1+m_2}{4m_1m_2}p^2
    \; ,\\
    h &=
    \frac{m_1-m_2}{4m_1m_2} p^2
    \; ,\\
    \n &=
    \frac{2p_xp_y \hat{x}+ (p_x^2-p_y^2)\hat{z}}{p^2}\; ,
\end{align}
with $\hat{x}$ and $\hat{z}$ the unit vectors in the momentum space $x,z$ directions. Correspondingly, we can determine all nonvanishing geometric quantities.

We begin by evaluating the metric $(g_{ij})$ and the Christoffel symbols $(\Gamma_{ijk})$:
\begin{align}
    \partial_{p_x} \n
    &=
    -2p_y\frac{(p_x^2-p_y^2)\hat{x}-2p_xp_y \hat{z}}{p^2}\; ,
    \nonumber\\
    \partial_{p_y} \n
    &=
    2p_x\frac{(p_x^2-p_y^2)\hat{x}-2p_xp_y \hat{z}}{p^2}\; ,
    \nonumber\\
    g_{xx}
    &=
    \frac{4p_y^2}{p^4}\;,
    \\
    g_{yy}
    &=
    \frac{4p_x^2}{p^4}\;,
    \\
    g_{xy}=g_{yx}
    &=
    -\frac{4p_xp_y}{p^4}\;,
\\
    \G_{xxx}
    &=
    \frac{1}{2}\partial_{p_x}g_{xx}
    =
    -\frac{8p_xp_y^2}{p^6}\; ,
    \\
    \G_{yyy}
    &=
    \frac{1}{2}\partial_{p_y}g_{yy}
    =
    -\frac{8p_x^2p_y}{p^6}\; ,
    \\
    \G_{xxy}=\G_{xyx}
    &=
    \frac{1}{2}(\partial_{p_y}g_{xx}+\partial_{p_x}g_{xy}-\partial_{p_x}g_{xy})
    =
    \frac{4p_y(p_x^2-p_y^2)}{p^6}\; ,
    \\
    \G_{yyx}=\G_{yxy}
    &=
    \frac{1}{2}(\partial_{p_x}g_{yy}+\partial_{p_y}g_{yx}-\partial_{p_y}g_{xy})
    =
    -\frac{4p_x(p_x^2-p_y^2)}{p^6}\; ,
    \\
    \G_{xyy}
    &=
    \frac{1}{2}(\partial_{p_y}g_{xy}+\partial_{p_y}g_{xy}-\partial_{p_x}g_{yy})
    =
    \frac{8p_xp_y^2}{p^6}\;,
    \\
    \G_{yxx}
    &=
    \frac{1}{2}(\partial_{p_x}g_{yx}+\partial_{p_x}g_{yx}-\partial_{p_y}g_{xx})
    =
    \frac{8p_x^2p_y}{p^6}\;,
\end{align}
from which we can further obtain $F_{ij}$:
\begin{align}
    F_{xx}&=F_{yy}=F_{xy}=F_{zx}=F_{zy}
    = 0
    \;,\\
    F_{xz}
    &=
    -\frac{g_{xk}\varepsilon_{klz}\partial_{p_l}h_0}{4h}
    -\frac{1}{8}\varepsilon_{zkl}\G_{lkx}
    \nonumber\\
    &=-\frac{g_{xx}\varepsilon_{xyz}\partial_{p_y}h_0}{4h}
    -\frac{g_{xy}\varepsilon_{yxz}\partial_{p_x}h_0}{4h}
    -\frac{1}{8}\varepsilon_{zxy}\G_{yxx}
    -\frac{1}{8}\varepsilon_{zyx}\G_{xyx}
    \nonumber\\
    &=-\frac{p_y}{2p^2}\frac{m_1+m_2}{m_1-m_2}\frac{4p_y^2}{p^4}
    -\frac{p_x}{2p^2}\frac{m_1+m_2}{m_1-m_2}\frac{4p_xp_y}{p^4}
    -\frac{1}{8}\frac{8p_x^2p_y}{p^6}
    +\frac{1}{8}\frac{4p_y(p_x^2-p_y^2)}{p^6}
    \nonumber\\
    &=
    -\frac{2p_y}{p^4}\frac{m_1+m_2}{m_1-m_2}
    -\frac{p_y}{2p^4}
    \nonumber\\
    &=
    -\frac{p_y}{2p^4}
    \lb \frac{1}{4}+\frac{m_1+m_2}{m_1-m_2}\rb\; ,
    \\
    F_{yz}
    &=
    -\frac{g_{yk}\varepsilon_{klz}\partial_{p_l}h_0}{4h}
    -\frac{1}{8}\varepsilon_{zkl}\G_{lky}
    \nonumber\\
    &=-\frac{g_{yx}\varepsilon_{xyz}\partial_{p_y}h_0}{4h}
    -\frac{g_{yy}\varepsilon_{yxz}\partial_{p_x}h_0}{4h}
    -\frac{1}{8}\varepsilon_{zxy}\G_{yxy}
    -\frac{1}{8}\varepsilon_{zyx}\G_{xyy}
    \nonumber\\
    &=\frac{p_y}{2p^2}\frac{m_1+m_2}{m_1-m_2}\frac{4p_xp_y}{p^4}
    +\frac{p_x}{2p^2}\frac{m_1+m_2}{m_1-m_2}\frac{4p_x^2}{p^4}
    +\frac{1}{8}\frac{4p_x(p_x^2-p_y^2)}{p^6}
    +\frac{1}{8}\frac{8p_xp_y^2}{p^6}
    \nonumber\\
    &=\frac{2p_x}{p^4}\frac{m_1+m_2}{m_1-m_2}
    +\frac{p_x}{2p^4}
    \nonumber\\
    &=
    \frac{p_x}{2p^4}
    \lb \frac{1}{4}+\frac{m_1+m_2}{m_1-m_2}\rb\; .
\end{align}
An analogous result for the other band ($n=2$) can be obtained by swapping $m_1$ and $m_2$. Furthermore, we can calculate the nonlinear conductivity tensor $\chi^\text{orb}_{ij,k}$ defined by $j_i=\chi^\text{orb}_{ij,k}E_j B_k$, namely (note that the indexing is different in Refs.~\cite{Gao2014, Wang2024_1, Wang2024_2}):

\begin{equation}
    \chi^\text{orb}_{ij,k}
    =
    -\int[d\pv]\frac{\partial f_1}{\partial \ep_1}(\partial_{p_i}\ep_1 F_{jk}-\partial_{p_j}\ep_1 F_{ik}),
\end{equation}
and the nonvanishing term is given by:
\begin{align}
    \chi^\text{orb}_{xy,z}
    &=
    -\int[d\pv]\frac{\partial f_1}{\partial \ep_1}
    (\partial_{p_x}\ep_1 F_{yz}-\partial_{p_y}\ep_1 F_{xz})
    \nonumber\\
    &=-\frac{1}{m_1}\int[d\pv] \frac{\partial f_1}{\partial \ep_1} \frac{1}{2p^2}
    \lb \frac{1}{4}+\frac{m_1+m_2}{m_1-m_2}\rb
    \nonumber\\
    &\sim
    -\frac{1}{4\pi^2}\frac{1}{m_1}\int d\ep_1 m_1 \frac{\partial f_1}{\partial \ep_1}
    \frac{1}{4m_1\ep_1}
    \lb \frac{1}{4}+\frac{m_1+m_2}{m_1-m_2}\rb.
\end{align}
At zero temperature ($T=0$), we find:
\begin{align}
    \chi^\text{orb}_{xy,z}
    &\sim
    -\frac{1}{4\pi^2}\frac{1}{m_1}\int d\ep_1 m_1 \d(\ep_1-\mu)
    \frac{1}{4m_1\ep_1}
    \lb \frac{1}{4}+\frac{m_1+m_2}{m_1-m_2}\rb
    \nonumber\\
    &=-\frac{1}{16\pi^2}\frac{1}{m_1\mu}\lb \frac{1}{4}+\frac{m_1+m_2}{m_1-m_2}\rb.
\end{align}
In the limit of one of the bands being flat ($m_2 \rightarrow \infty$), we simply obtain:
\begin{equation}\label{chi_flat}
    \chi^\text{orb}_{xy,z}
    =\frac{3}{64\pi^2}\frac{1}{m_1\mu},
\end{equation}
and if both bands contribute, we instead have:
\begin{align}
    \chi^\text{orb}_{xy,z}
    &= 
    -\frac{1}{16\pi^2}\frac{1}{m_1\mu}\lb \frac{1}{4}+\frac{m_1+m_2}{m_1-m_2}\rb
    +(m_1\leftrightarrow m_2)
    \nonumber\\
    &=
    \frac{3}{64\pi^2}\frac{1}{\mu}\frac{m_1+m_2}{m_1m_2}.
\end{align}
Numerically, we verify these analytical predictions against the three-band kagome model:
\beq{tb_kag}
   H(k_x, k_y)
    =
    \begin{pmatrix}
        0&t_0\cos{\frac{k_x}{2}}& t_0\cos{\frac{k_x-\sqrt{3}k_y}{4}}\\
        t_0\cos{\frac{k_x}{2}}& 0 & t_0\cos{\frac{k_x+\sqrt{3}k_y}{4}}\\
        t_0\cos{\frac{k_x-\sqrt{3}k_y}{4}}
        & t_0\cos{\frac{k_x+\sqrt{3}k_y}{4}} & 0
    \end{pmatrix}
\eec
consistent with the momentum-space representations of the general class of tight-binding Hamiltonians provided in the real-space orbital basis in Methods.\\

\begin{figure}[h]
    \centering
    \includegraphics[width=0.8\linewidth]{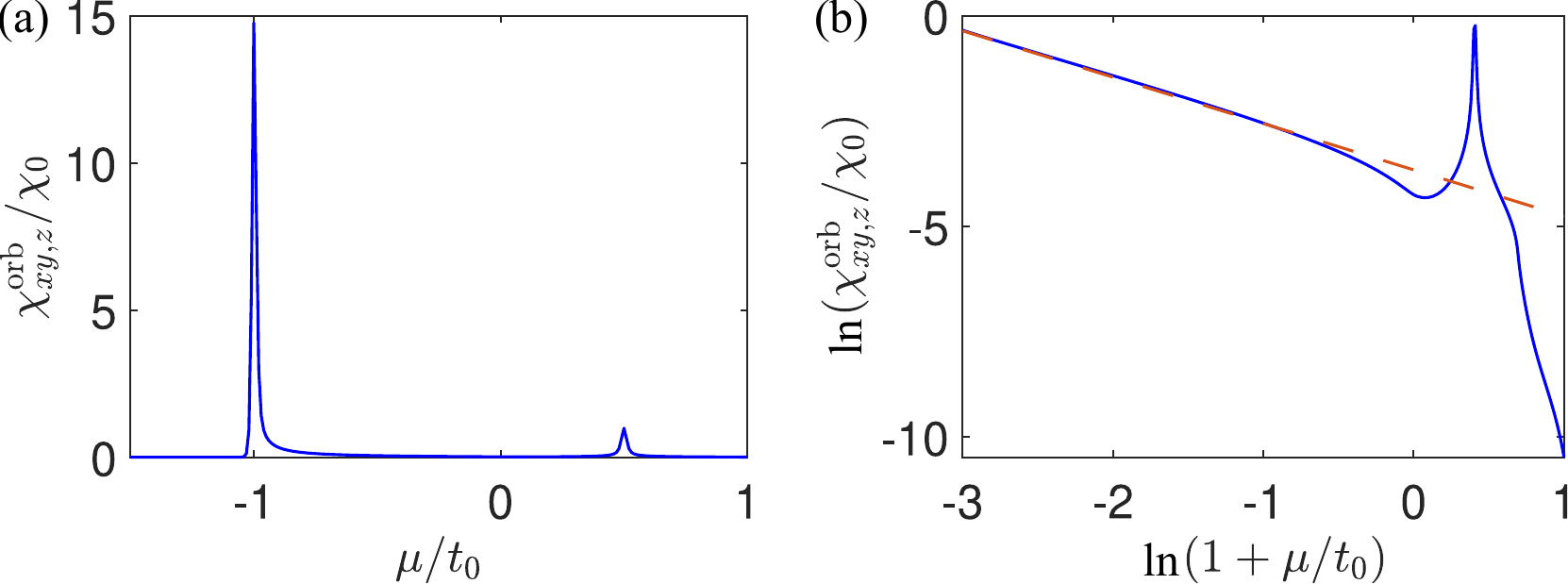}
    \caption{Numerical result for the orbital magnetononlinear Hall effect tensor $\chi^\text{orb}_{xy,z}$ for kagome Hamiltonian defined in Eq.~\eqref{tb_kag}, at finite temperature $T=10^{-3}t_0$. {\bf (a)} An Euler node with $e_2 = 1$ is present at $\mu=-t_0$ in the model, leading to an anomalous contribution to the nonlinear conductivity tensor. {\bf (b)} Log-log scaling of the response tensor $\chi^\mathrm{orb}_{xy,z}$,  showcasing that $\chi^\mathrm{orb}_{xy,z}$ scales as $(1+\mu/t_0)^{-1.1}$ within the numerical result, closely to the continuum model prediction of the $(1+\mu/t_0)^{-1}$ scaling in Eq.~\eqref{chi_flat}. }
    \label{fig:S1}
\end{figure}
\bibliography{./references.bib}